%% file: deboer_dm08.tex
\documentclass[final]{aipproc}
\layoutstyle{6x9}
\usepackage{graphicx}
\usepackage{natbib}
\bibpunct{(}{)}{;}{a}{}{,}
\usepackage{graphicx,amsmath}

\begin{document}

\title{Indirect Dark Matter Signals from EGRET and PAMELA   compared}


\author{Wim de Boer}{
  address={Physikhochhaus, IEKP, Universit\"at Karlsruhe\\
Postfach 6980, D-76128 Karlsruhe\\
Germany\\
e-mail: wim.de.boer@cern.ch}
}
\input{config.tex}

\begin{abstract}

Dark Matter annihilation (DMA)
may yield an excess  of
gamma rays and antimatter particles, like antiprotons and positrons,
above the background from cosmic ray interactions.
The excess of diffuse Galactic Gamma Rays from EGRET shows all the features expected from DMA.
The new precise measurements of the antiproton and positron fractions from PAMELA are compared with
the EGRET excess. It is shown that the charged particles are strongly dependent on the propagation model used.
The usual propagation models with isotropic propagation models are incompatible with the recently observed
convection in our Galaxy. Convection leads to an order of magnitude uncertainty in the yield of charged particles from DMA, since even a rather small convection will let drift the charged particles in the halo to outer space. It is shown that  such anisotropic propagation models including convection
 prefer a contribution from DMA for the antiprotons, but the rise in the positron fraction, as observed by PAMELA, is incompatible with  DMA, if compared with the EGRET excess.   A rise in the positron/electron ratio is expected, if the observed rise in the proton/electron ratio is carefully fitted in a propagation model, although the rise is slightly larger than expected, so additional local sources may contribute as well within the limited accuracy of the data.

\end{abstract}

\maketitle

\section{Introduction}
Present cosmological data has shown that DM makes up 23\% of the energy of the universe as compared to 4.4\% for the known baryonic matter \cite{wmap}. Candidates for DM are e.g. the lightest supersymmetric particles, which are stable and weakly interacting with normal matter, thus explaining why the DM forms large haloes around the cores of visible matter inside the galaxies.  If DM is  a thermal relic from the early universe, its low number density compared to the photon density can be explained, if they annihilated with each other into normal matter, just like the protons disappeared by annihilation with antiprotons. However, the annihilation rate of protons is so high compared with the expansion rate of the universe, that one has to invoke an excess of protons over antiprotons to explain the remaining protons, which implies CP-violation. DM particles are neutral and therefore  likely their own antiparticles, so the annihilation cannot stop because of CP-violation. Therefore a remaining amount of a neutral thermal relic can only be explained if the expansion rate of the universe is comparable or smaller than the annihilation rate. This gives a direct relation between the amount of DM, the Hubble constant and the annihilation rate, which yields \cite{griest}: $<\sigma v>= 3\cdot 10^{-27}/\Omega_{DM}h^2=3\cdot10^{-26}~cm^3/s$. For  typical velocities of $10^{-3}$c of cold DM particles this yields an annihilation cross section of the order of $10^{-33}~cm^2$, practically independent of the mass of the DM particle.
Note that this annihilation cross section is at least 10 orders of magnitude larger than the spin-independent elastic scattering cross section of DM particles on a proton or neutron, as is known from the limits from direct DM searches. Surprisingly, both the annihilation cross section and the scattering cross section are  in the ballpark expected for the lightest
supersymmetric particles, the neutralinos.

Neutralinos annihilate into mono-energetic quarks, which hadronize mainly into light charged and neutral pions.
The neutral pions decay almost exclusively into gamma rays with a characteristic spectrum  known from high energy $e^+e^-$ annihilation experiments \cite{pdb}. The observed EGRET excess of diffuse Galactic gamma rays shows exactly such a characteristic spectrum in all sky directions \cite{us,wdb1,gebauer}
 and is perfectly consistent with a 50-100 GeV neutralino \cite{egret_susy}.

Recently published data from the PAMELA satellite experiment on the antiproton  fluxes  \cite{pamela_pb} and the
positron fraction \cite{pamela_ep}, defined as the ratio of fluxes of positrons and the sum of electrons and positrons, i.e. $e^+/(e^+ + e^-)$, has drawn large interest in the community interested in dark matter annihilation (DMA) signals from our Galaxy, see e.g. Refs \cite{pamela_response}. Usually
the antiproton data is considered to be standard, i.e. in agreement with so-called isotropic propagation models, while the increase in the positron
fraction disagrees with predictions from such models.
Also the electron spectrum is claimed to show an anomaly
in the range of 200-600 GeV \cite{atic,ppb-bets}.
Having no excess in antiprotons but a large excess in positrons or electrons is not expected from neutralino annihilation, so alternatives, like local sources, e.g. pulsars, \cite{pulsars} or other DM candidates  (see e.g. \cite{axions}) have been considered as well.

Anomalies implies one compares the data with some model about the cosmic ray fluxes and the propagation  in our Galaxy. Here one usually assumes  simple isotropic propagation, as implemented in the publicly available GALPROP program \cite{galprop}.
However, isotropic propagation models are basically ruled out with the observation of convective streams deduced from the X-ray spectra of the gas
in the halo by the ROSAT satellite \cite{rosat, breitschwerdt_nature}. Convection lets drift the cosmic rays (CRs) in the halo to outer space,  which strongly reduces the positron and antiproton DMA signals in anisotropic propagation models \cite{gebauer}.

 In this paper we compare the EGRET excess of gamma rays with the PAMELA data using anisotropic propagation models and find that the interpretation of the EGRET excess as a signal of DMA is
 perfectly consistent with the PAMELA data: the DMA shows the expected contribution to
 the antiproton flux, while the DMA contribution to positrons is negligible.

\begin{figure}
{\includegraphics[width=0.346\textwidth,height=0.3\textwidth,clip]{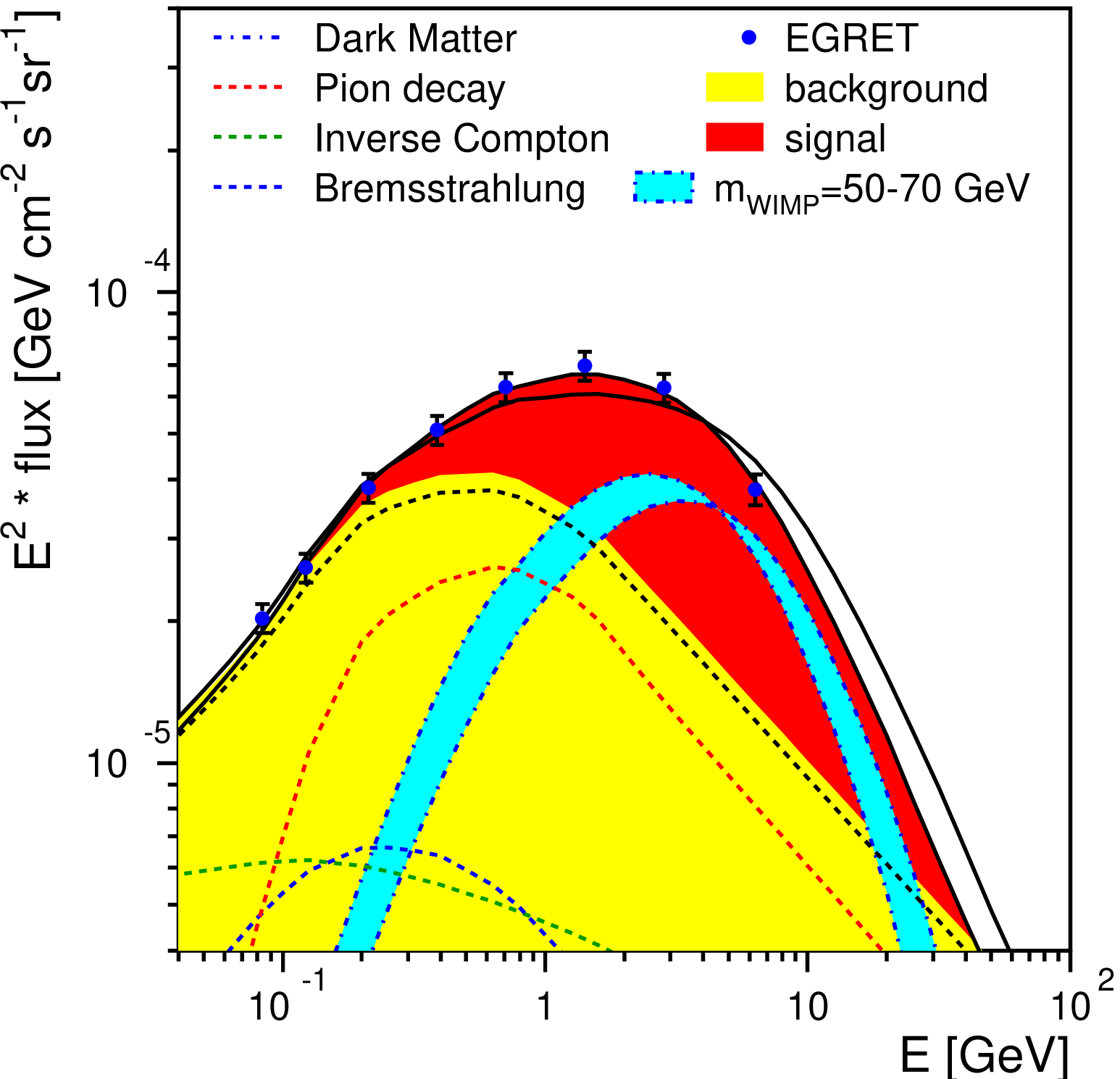}}
{\includegraphics[width=0.346\textwidth,height=0.3\textwidth,clip]{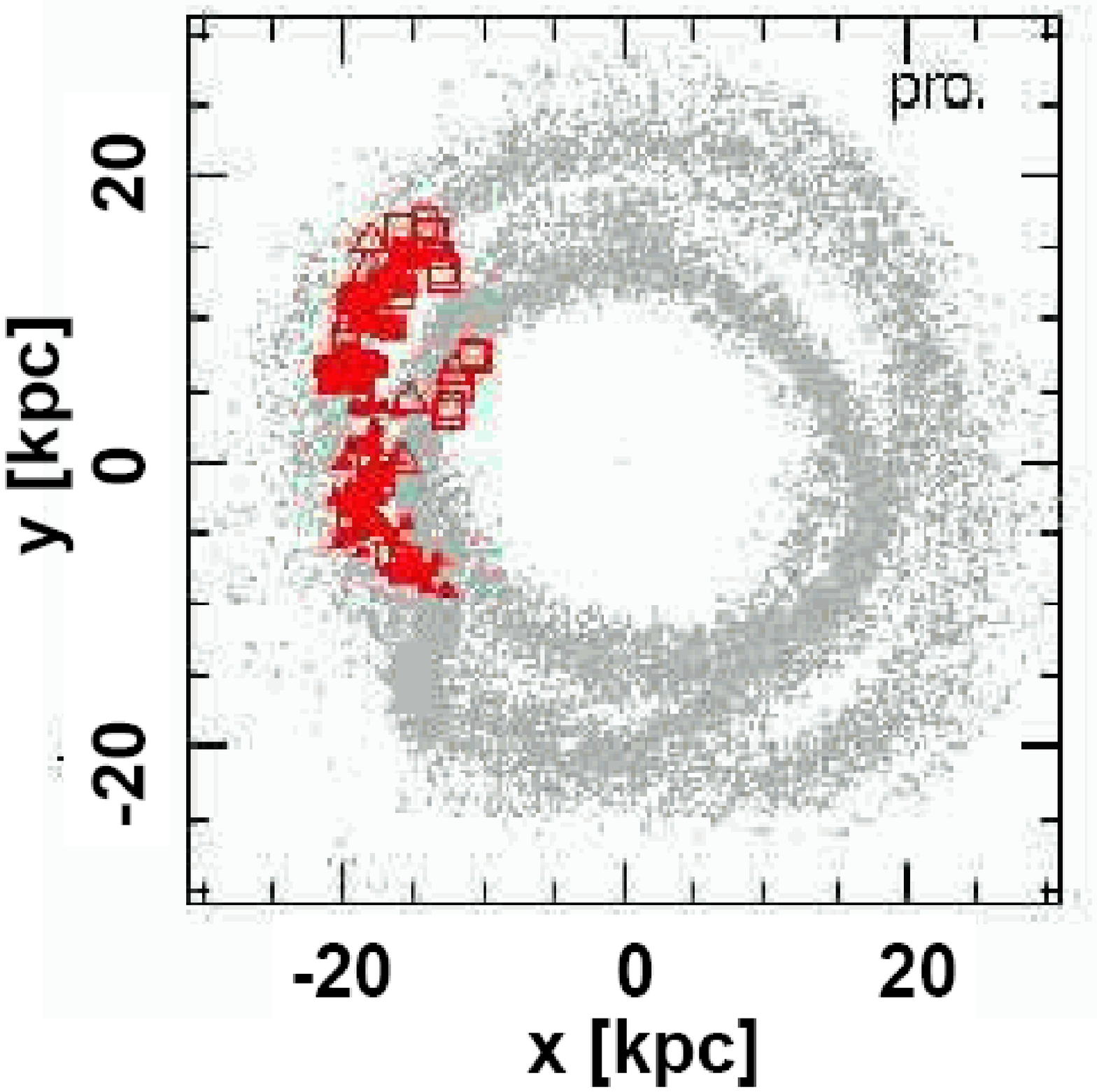}}
{\includegraphics[width=0.346\textwidth,height=0.33\textwidth,clip]{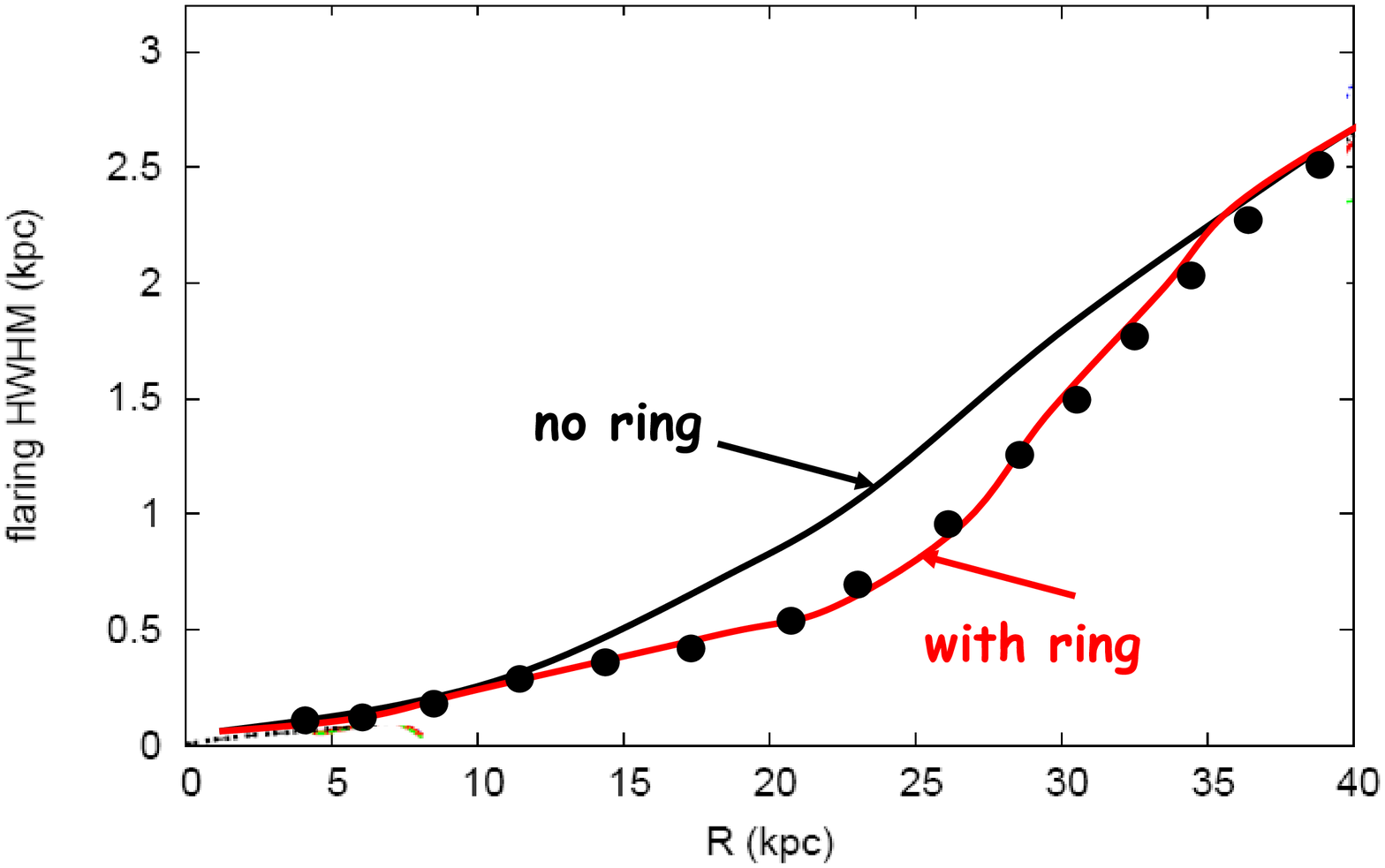}}
\caption{Left: Fit of the shapes of background and DMA signal to the EGRET data in the direction of the Galactic center. The light shaded (yellow) area indicates the background using the shapes known from accelerator experiments, while the dark shaded (red) area corresponds to the signal contribution from DMA for a 60 GeV WIMP mass, where the small intermediate (blue) shaded area corresponds to a variation of the WIMP mass between 50 and 70 GeV. Centre:  Results of an N-body simulation of the tidal disruption of the
Canis Major dwarf Galaxy, whose orbit was fitted to the observed stars (red data points). The simulation predicts a ringlike structure of dark matter with a radius of 13 kpc. From \cite{penarrubia}.
Right: The half-width-half-maximum (HWHM) of the gas layer of the Galactic disk as function of the distance from the Galactic center. Clearly, the fit including a ring of dark matter above 10 kpc describes the data much better. Adapted from data in \cite{kalberla}.
\label{f1}}
\end{figure}
\section{The EGRET excess of diffuse Galactic gamma rays}\label{egret}

An excess of diffuse gamma rays  has
 been observed by the EGRET telescope on board of NASA's CGRO (Compton Gamma
Ray Observatory)\cite{hunter}. Below 1 GeV the cosmic ray (CR)
interactions describe the data  well, but above 1 GeV the data are up to a
factor two above the expected background. The excess shows all the features of DMA
 for a WIMP mass between 50 and 70 GeV, as shown in Fig. \ref{f1} \cite{us}.
In addition, the results are perfectly consistent with
the expectations from Supersymmetry\cite{egret_susy}.

The  analysis of the EGRET data was performed with a so-called data-driven calibration of the background,
a procedure commonly used in accelerator experiments to reduce the sensitivity to model dependence of signal and background calculations. Such analysis techniques are rather unusual in the astrophysics community or among theorists,
who typically use the standard procedure of taking a Galactic model to calculate the background and a
 certain DM halo model to calculate the signal and then compare signal plus background with data.
Such analysis are highly sensitive to uncertainties in the background  and DM halo models.

A data-driven approach is particularly suitable for the analysis of gamma rays, since the shape of the dominant background, which is the $\pi^0$ production in inelastic collisions of CR protons on the hydrogen gas of the disk, is well known from so-called fixed target accelerator experiments in which a proton beam is scattered on a hydrogen target \cite{pdb}.
Furthermore, the shape of the DMA signal is known from $e+e-$ annihilation, so the gamma ray shapes of both, signal and background are known from accelerator experiments with high precision, since these reactions happen to be the best studied ones in high energy physics \cite{pdb}. Since the signal has a significantly harder spectrum than the background one can perform a data-driven analysis by simply fitting the two shapes to the experimental data
with a free normalization for each shape,  thus obtaining the absolute contribution from signal and background for each sky direction in a rather model-independent way. Uncertainties in the interstellar background shape arise from solar modulation and in addition from the uncertainties from electron CRs generating gamma-rays by inverse Compton scattering and Bremsstrahlung. However, since the electron flux of CRs is two orders of magnitude below the proton flux, this effect can be included with sufficient accuracy.

A simultaneous fit of DMA signal and background shape has been made to 180 independent sky directions.
The average $\chi^2$ per degree of freedom summed over all ca. 1400 data points is around 1, indicating that the errors are correctly estimated. But above all such a good $\chi^2$ implies that the main conditions for a signal of DMA are fulfilled, namely i) the {\it shape} of the excess  corresponds to the fragmentation of mono-energetic quarks with the same energy in {\it all} sky directions and ii) it was found that the {\it intensity} distribution of the excess  agrees with the mass distribution, as deduced from the  rotation curve \cite{us}
iii) the background distribution agrees within errors with the expectation from  GALPROP, the most up-to-date Galactic propagation model \cite{galprop}, as can be seen from Fig. 3 in Ref. \cite{us}.

 The derived DM halo profile  shows some unexpected substructure: outside the disk it corresponds to a cored halo profile, but inside the disk it reveals two additional doughnut-like structures at distances of about 4 and 13 kpc from the Galactic center.  Structures are expected from the tidal disruption of dwarf galaxies captured in the gravitational field of our Galaxy.
The "ghostly" ring of stars or Monocerus stream (with about $10^8-10^9$ solar masses in visible matter) could be the tidal streams of the Canis Major dwarf galaxy (see e.g. \cite{cma,penarrubia} and references therein). If so, the tidal streams  predicted from N-body simulations are perfectly consistent with the ring at 13 kpc  \cite{penarrubia}, as shown in the central panel of Fig. \ref{f1}.  The strong gravitational potential well in this stream was  confirmed from the gas flaring, which is reduced at the position of the ring \cite{kalberla}.
The half-width-half-maximum of the gas layer in the disk is shown on the right-hand panel of Fig. \ref{f1}. The reduced gas flaring corresponds to more than $10^{10}$ solar masses, in  agreement with the EGRET ring. It should be noted that the peculiar shape of the gas flaring was only understood after the astronomers heard about the EGRET ring. The effect is so large that visible matter cannot explain this peculiar shape, simply because there is not much visible matter above 10 kpc. Also the peculiar change in the slope of the rotation curve at a Galactic radius around 10 kpc can only be explained by a ringlike structure \cite{us}. A similar ring
in the outer disk has been discovered in a nearby galaxy, indicating that such infalls may shape
the disk and its warps \cite{penarrubia1}.

So the DMA interpretation of the EGRET excess at 13 kpc is strongly supported by these independent astronomical observations. The ring at 4 kpc might also originate from the disruption of a smaller dwarf galaxy, but here the density of stars is too high to find evidence for tidal streams. However, direct evidence of a stronger gravitational potential well in this region comes from the ring of dust at this location. Since this ring is slightly tilted with respect to the plane its presence and orientation are can be explained by the presence of a ringlike structure of DM.
It should be noted that such substructures are most easily discovered by a   data-driven approach,
which does not rely on propagation and DM halo models.

\section{Antiprotons in anisotropic propagation models}
\begin{figure}
\includegraphics[width=0.35\textwidth,height=0.32\textwidth,clip]{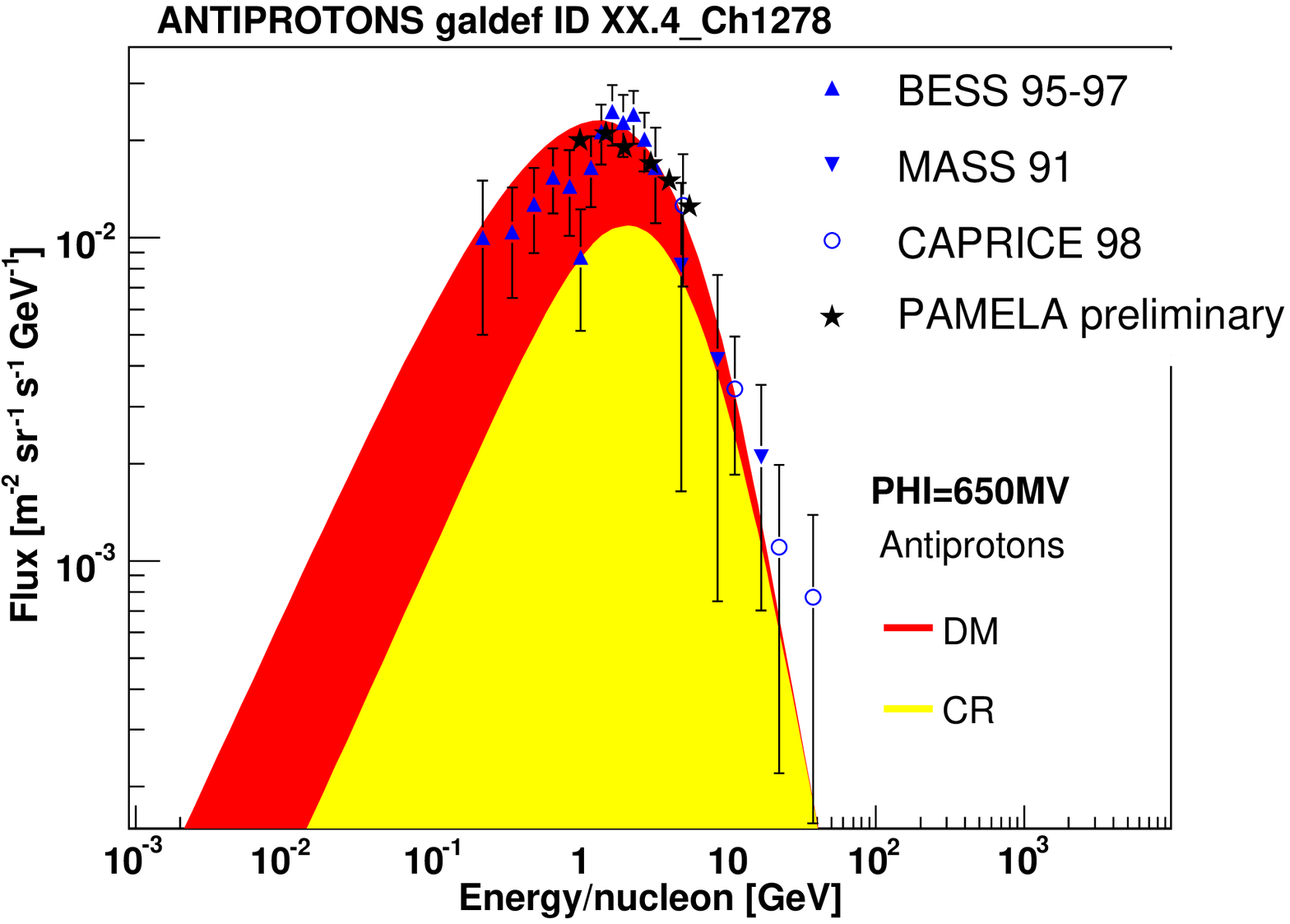}
\includegraphics[width=0.32\textwidth,height=0.32\textwidth,clip]{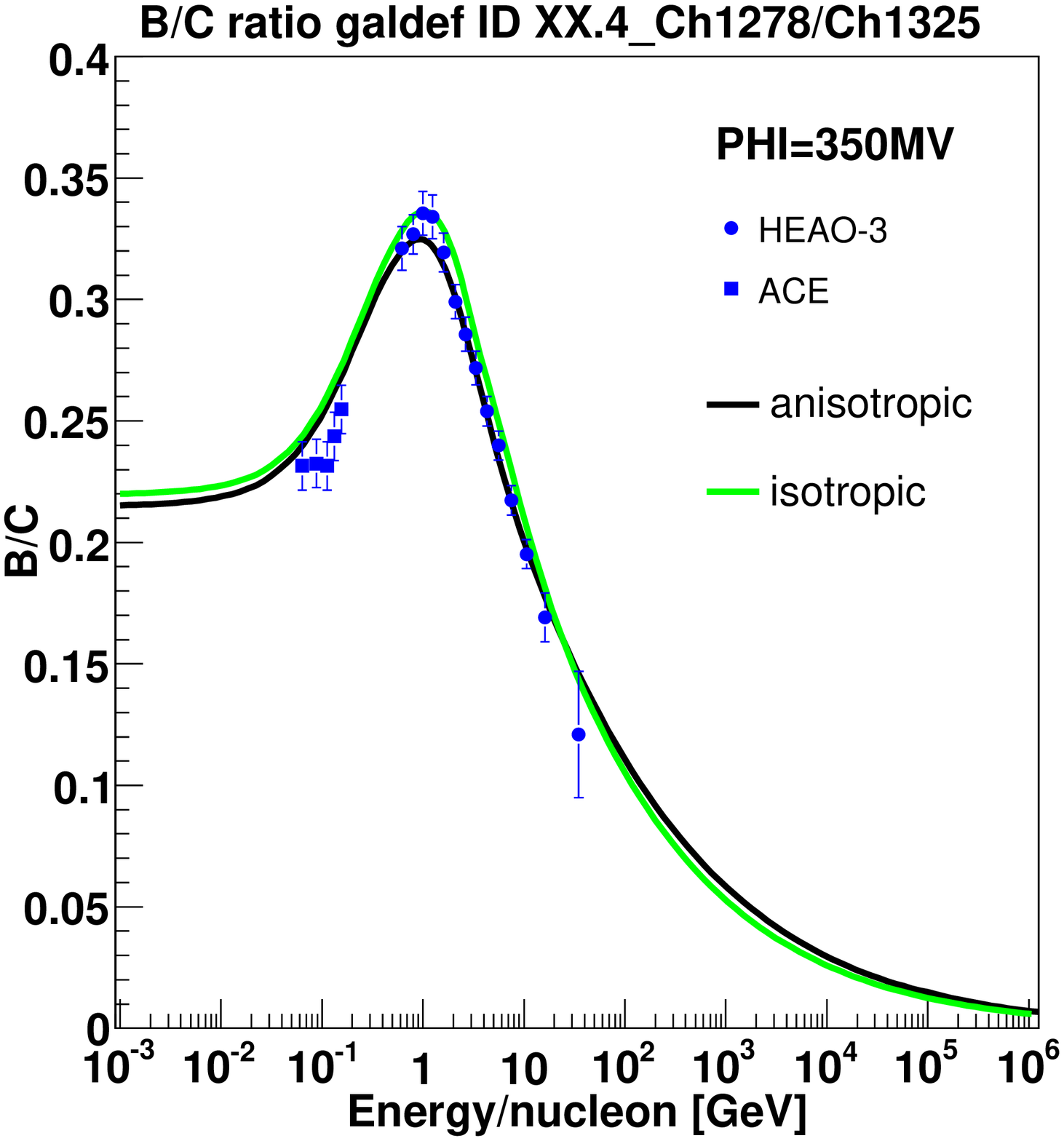}
\includegraphics[width=0.32\textwidth,height=0.32\textwidth,clip]{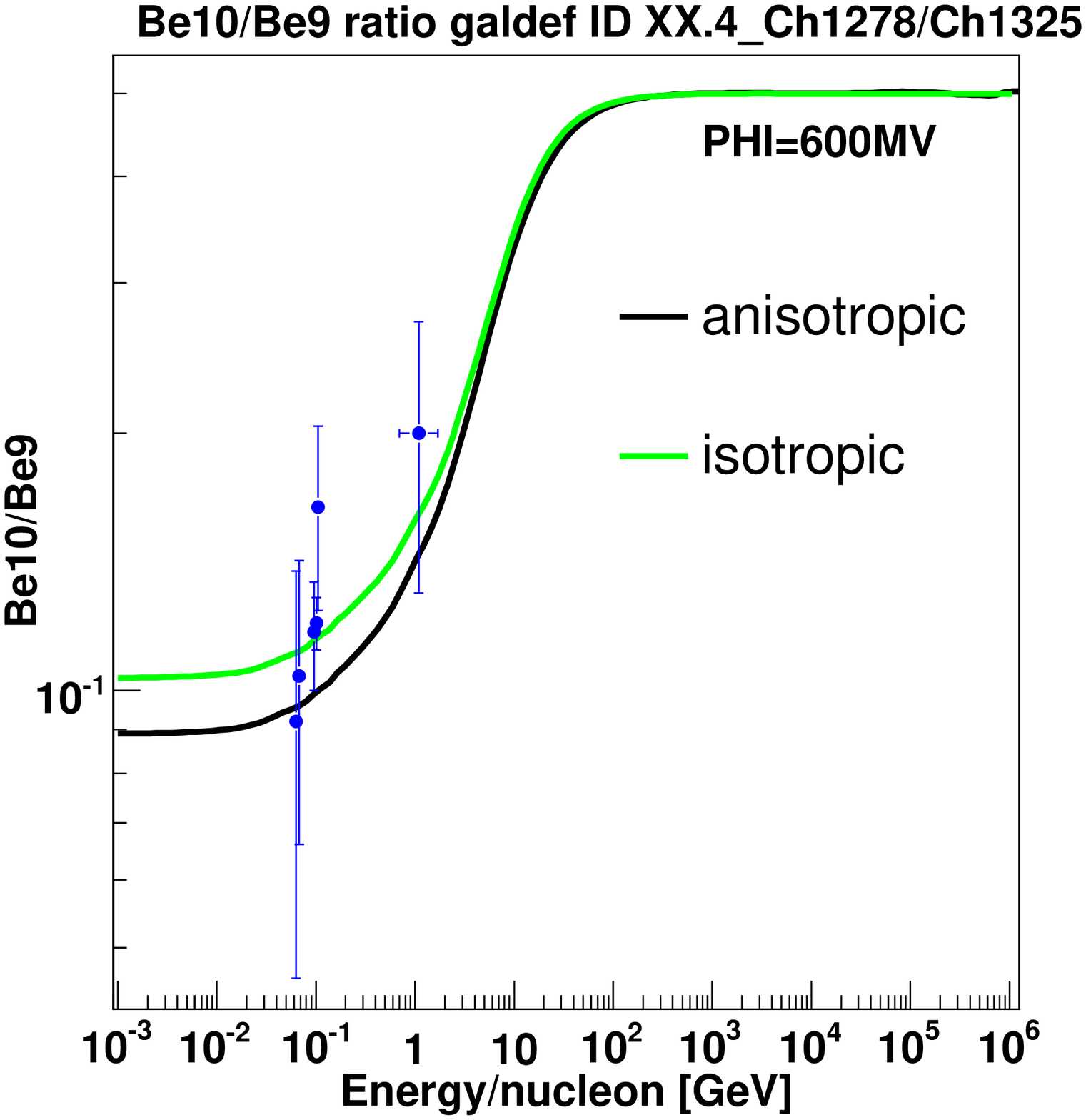}
\caption{
Data on antiprotons, $B/C$ and $^{10}Be/^9Be$ compared with propagation models including DMA.}
\label{f2}
\end{figure}
 High energy cosmic rays (CRs) travel close to the speed of light. Nevertheless,
their averaged speed is much reduced by the interaction with the plasma created by the
ensemble of CRs in the Galaxy. The resulting propagation is usually described as a diffusion process
combined with a transport perpendicular to the disk by Galactic winds originating from
Supernovae (SN) explosions. Details about basic propagation principles  can be found in a recent review \cite{strong_rev}.
%

Isotropic propagation models cannot include convection speeds above ca. 10 km/s, since in these models secondary particles, like Boron, are produced by the fragmentation of heavier nuclei (mainly C, N, O) hitting the gas of the target. Therefore too high convection lets drift the CRs away from the disk, thus producing too few secondaries.
However, the convection speeds from the ROSAT data are significantly higher than the limit allowed by isotropic propagation models. Independent support for rather high convection comes from the large large bulge/disk ratio of the positron 511 keV annihilation line \cite{prantzos}. In propagation models without convection these low energy positrons annihilate near their source, since diffusion, which is proportional to the energy to some power, is practically absent for MeV particles. However, convection is independent of energy and these particles can be convected away easily to the halo, where they find no electrons to annihilate. The transport to the halo is additionally facilitated by the turbulent SN explosions, which generate high pressure bubbles blowing material out of the Galactic disk into chimneys reaching far into the halo \cite{breitschwerdt_SN}.

If  convection speeds above 150 km/s are included, the lifetime of CRs and the production of secondaries can be tuned to fit the data by adjusting the diffusion coefficient in the disk (in R-direction) and the halo (in z-direction) {\it independently}, thus leading to anisotropic propagation.  The diffusion in R has to be tuned such, that the transport from the source to us lasts at least $10^7$ years, as required by the "cosmic clocks", like the $^{10}Be/^9Be$ ratio. Given the different amount of matter and magnetic turbulence in the disk and the halo it is not unreasonable to have different diffusion speeds in R and z (cylindrical coordinates).
For example, the molecular cloud complexes (MCCs) in the disk may be efficient traps of CRs,
because of their high magnetic fields, which would trap CRs magnetically between MCCs, just as the magnetic field of the earth traps CRs in the van Allen belts. Such trapping would nicely explain why the low energy positrons from radioactive decays always annihilate with electrons from hydrogen atoms, not from molecules, as observed by INTEGRAL \cite{integral}, since the positrons would simply be reflected by the MCCs.
Note that  trapping can increase locally the CR density significantly. The large average distance between MCs allows to trap particles easily up to TeV energies and cause efficient reacceleration \cite{zirakashvili}. Trapping also improves the isotropy of CRs,  because  the surrounding traps scatter the CRs in all directions, as discussed in detail by \cite{chandran}. Such a trapping picture combined with convection can  reproduce i) the secondary production, ii) the small radial gradient in the production of gamma rays (because the CRs near the source are driven away by the Galactic winds from the disk, thus reducing the strong gamma ray production towards the Galactic center \cite{breitschwerdt_gammas}), iii) the large bulge/disk ratio for the positron 511 keV annihilation line \cite{prantzos}, because the positrons in the disk are simply transported to the halo by convection, where there are no electrons bound on atoms to annihilate and iv) the evidence for Galactic winds   from the ROSAT satellite \cite{rosat,breitschwerdt_nature}. More details on such anisotropic convection driven models (aCDM) will be published elsewhere \cite{gebauer_prop}.

The antiproton flux from nuclear interactions tends to be on the low side compared with the data from PAMELA\cite{pamela_pb}, CAPRICE \cite{caprice} and BESS \cite{bess}, as shown in Fig. \ref{f2}, but this can be nicely remedied by the antiproton flux from DMA, which has a similar shape as the background for WIMP masses in the order of 100 GeV.
The boost factor and the WIMP mass were taken from the EGRET analysis discussed above.
Note that the relative production of antiprotons and gamma rays is known from the fragmentation of quarks, as measured in $e^+e^-$ annihilation. Previous claims that the DMA interpretation is excluded by the too high antiproton fluxes from DMA \cite{bergstrom_pb} is not valid, if convection is included in the propagation model.
It should be reminded that propagation models without the DMA contribution to antiprotons
are challenged by the rather high antiproton flux \cite{moskalenko_pb}.
 A higher antiproton flux can be obtained, if one resorts to so-called
 optimized models, in which the global CR spectra are different from the local ones.
 Such optimized models have been devised to explain the EGRET excess without DMA \cite{om}.

 Plots concerning the $B/C$ and the $^{10}Be/^9 Be$ ratios are shown in Fig. \ref{f2} as well. Boron is a secondary particle produced mainly in the spallation of heavier primary particles, like C,N and O during collisions with the gas of the disk. Therefore, the ratio is sensitive to the gas distribution, the primary source distribution and the propagation. $^{10}Be$ is a radioactive nucleus with a decay time of $1.6\cdot 10^6$ years, so if the lifetime of a CR inside our Galaxy is longer than the decay time, one expects a significant reduction of $^{10}Be$ as compared to the stable $^{9}Be$, at least at low energies, where the  relativistic increase of the lifetime is small. As can be seen from Fig. \ref{f2} the secondary production and the transport time between source and our local solar system can be described in both - isotropic and anisotropic -
 propagation models.

\section{Positron fraction}
\begin{figure}
\includegraphics[width=0.32\textwidth,height=0.32\textwidth,angle=0]{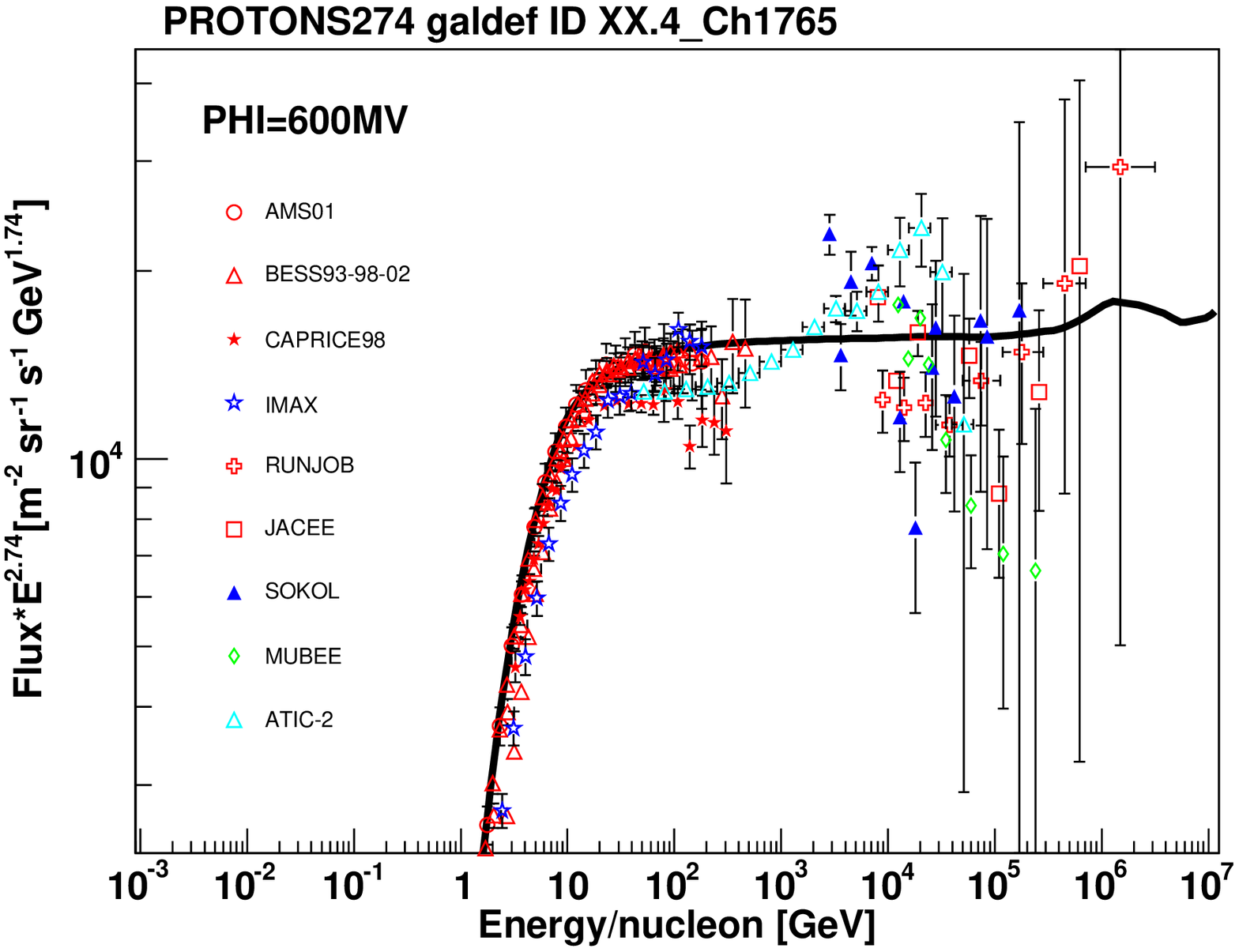}
\includegraphics[width=0.32\textwidth,height=0.32\textwidth,angle=0]{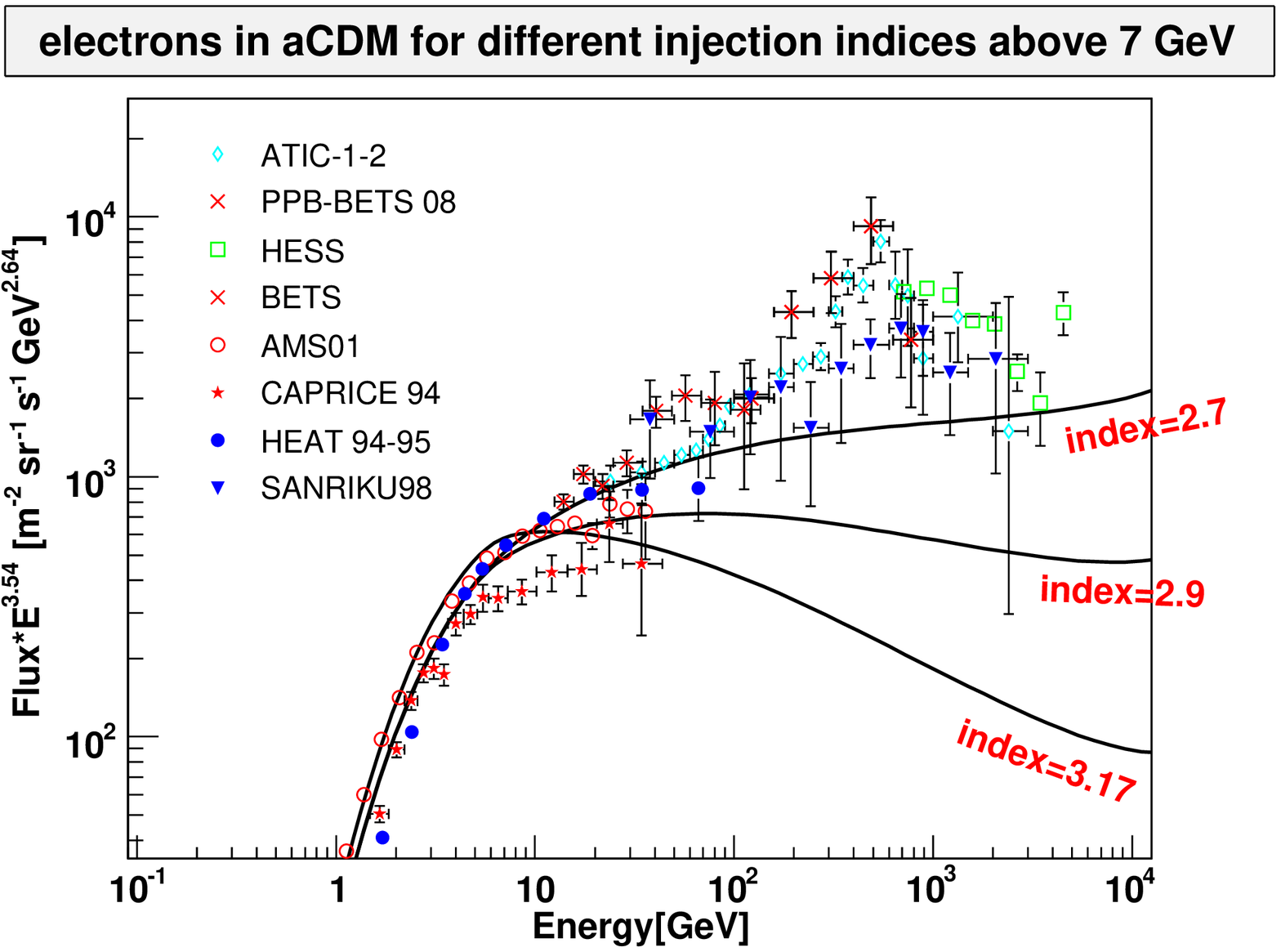}
\includegraphics[width=0.32\textwidth,height=0.32\textwidth,angle=0]{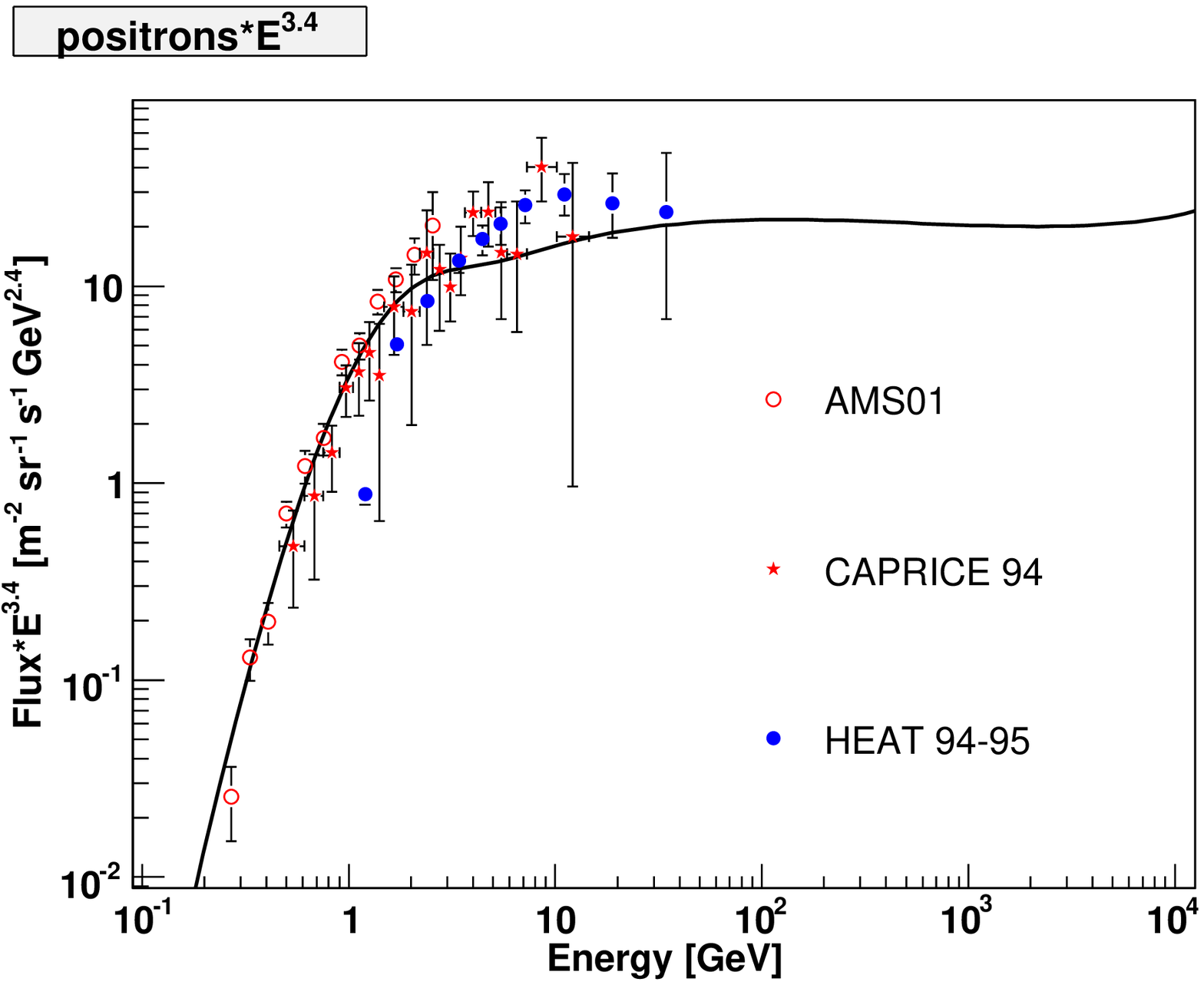}
\caption{The spectra of protons, electrons and positrons multiplied with a power of the
energy, as indicated on the vertical axis.
Data are taken from Ref. \cite{strong}.}
\label{f3}
\end{figure}
\begin{figure}
\includegraphics[width=0.45\textwidth,height=0.4\textwidth,angle=0]{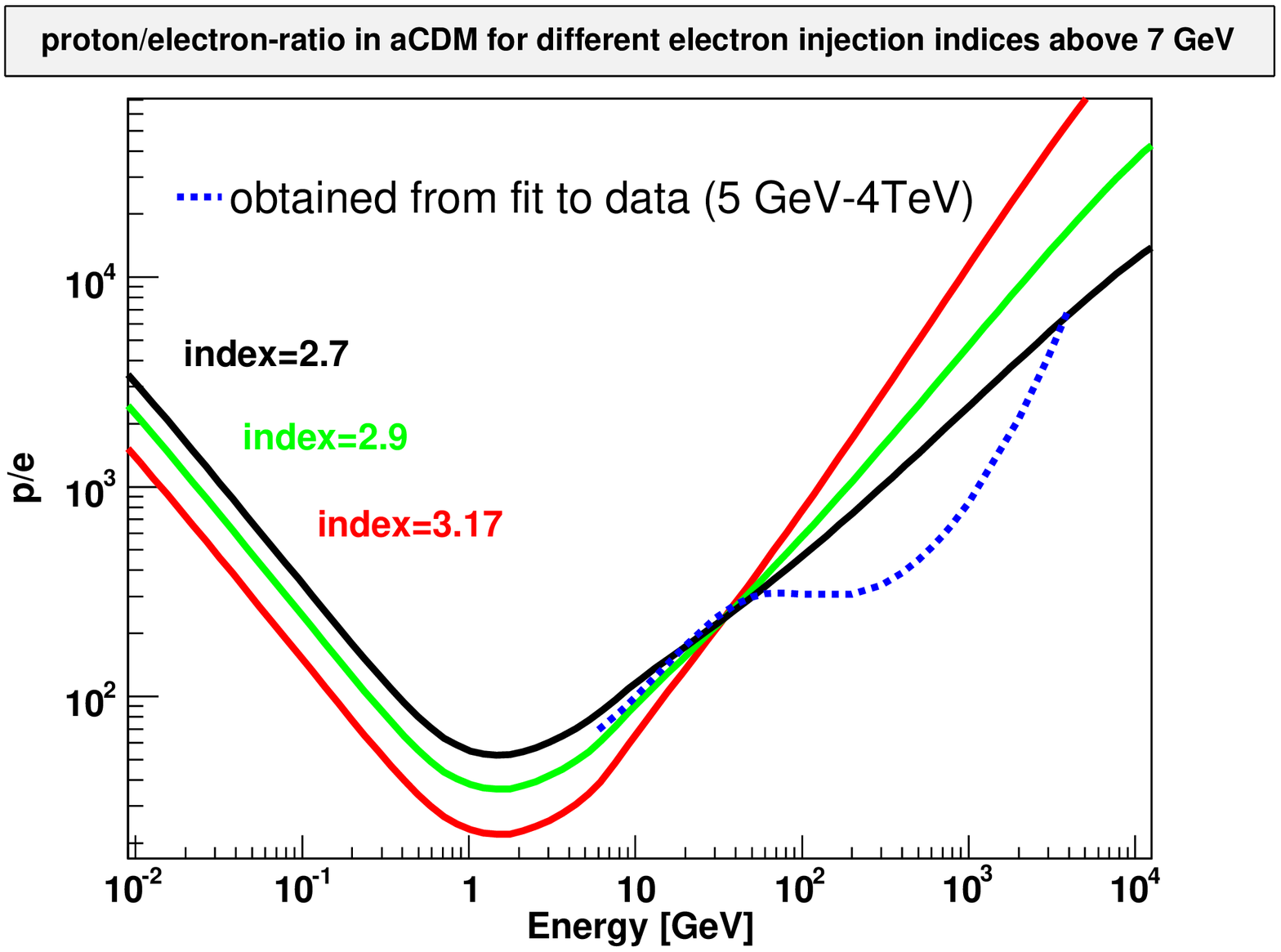}
\includegraphics[width=0.45\textwidth,height=0.4\textwidth,angle=0]{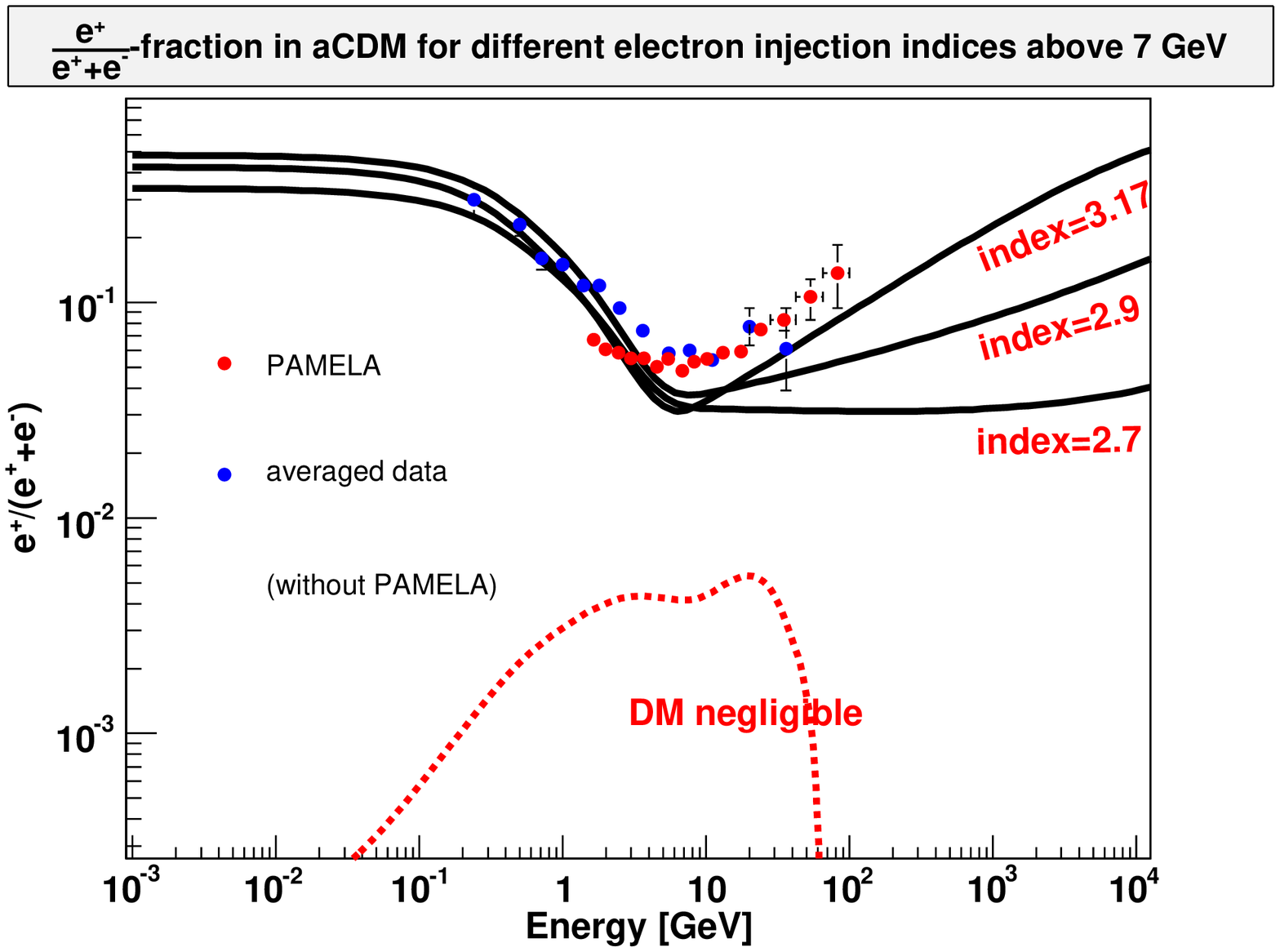}
\caption{Left: the ratio of proton and electron fluxes as function of energy.
Right: the positron fraction.
 The averaged HEAT and AMS-01 data are taken from Ref.~\cite{olzem}) and the new more precise PAMELA data are from Ref. \cite{pamela_ep}.
 The different solid curves correspond to different  injection indices of electrons, which are not yet very well constrained,  as shown in Fig. \ref{f3}.
 }
\label{f4}
\end{figure}
The positrons from DMA are produced mainly by the decays of positively charged pions produced after the hadronization of the quarks. The background comes from positive pions produced by inelastic collisions of CRs with the gas in the disk. Electrons are also produced by the decays of the negative pions produced in the same process, but this is only a small fraction compared to the electrons from SNRs. Unfortunately the spectra
of protons, electrons and positrons are not yet well measured, as shown in Fig. \ref{f3}.
Especially the more precise magnetic spectrometer experiments tend to show a power law
spectrum above a certain energy with a constant index, while the calorimetric experiments tend to show an excess above the extrapolation for this constant index. E.g. the HEAT and AMS-01 experiments are consistent with an injection index in the propagation model close to  -2.9, i.e. a power law $E^{-2.9}$.
 If one includes the calorimetric experiments at higher energies, a harder spectrum with an  index well above -2.7 is preferred.  This can be
due to local sources dominating above a certain energy, but without charge measurements of the particles the calorimetric experiments are prone to background from low energetic heavy nuclei. It is interesting to note that both the proton and electron spectra of ATIC show an increase above the index from the spectrometer experiments and then fall sharply at the edge of the acceptance. The increase in both,  protons and electrons, practically eliminates the DMA interpretation and strongly points to a local source, if it is not background from heavy nuclei increasing towards higher energies. The data have been taken from the Cosmic Ray Database by Strong, where the original references can be found \cite{strong}.

In Fig. \ref{f4} the ratio of protons/electrons is plotted for the best fitted electron data above 5 GeV (dotted curve), which is compared with the different injection indices
shown in the central panel of Fig. \ref{f3}. Independent of the electron injection index the p/e ratio shows an increase towards higher energy, as can be expected from  the higher energy losses of electrons towards higher energies. The positron spectrum is given by the proton spectrum modulo energy losses, so if the p/e ratio shows an increase, the positron spectrum may increase as well. This is indeed the case, as shown on the right hand panel for the three different injection indices. Clearly, the index of -2.9, as preferred from the magnetic spectrometer experiments, does not give a strong enough increase in the positron fraction, so a contribution from local sources is a likely possibility. Since our  local environment with the low gas density (so-called local bubble) was presumably created by a 10 to 20  recent supernova explosions \cite{breitschwerdt_SN}, nearby pulsars are expected.

The contribution from DMA is small, given the observed excess of Galactic gamma rays and taking the same boost factor  also for the positrons, as shown by the lower curve in the right hand panel of Fig. \ref{f4}. This is expected, since most of the positrons from DMA are produced in the halo, thus drifting away by convection, while the background is produced mainly in the disk.
For antiprotons the background is strongly reduced by the large threshold of CR protons, which requires protons  to be effectively above 10 GeV, while for the light pions there is hardly any threshold effect. Therefore the relative DMA contribution   is expected to be  larger for antiprotons than for positrons.

\section{Conclusion}\label{conclusion}
It is shown, that the interpretation of the EGRET excess as a signal of DMA is perfectly consistent with the new PAMELA data on positrons and antiprotons, if one uses a propagation model including the recently deduced convection from the X-ray spectra observed with the ROSAT satellite \cite{breitschwerdt_nature}. This leads to anisotropic propagation, in which the charged particles from the halo drift  to outer space and the observed fluxes of charged
particles are largely produced locally.  The consequence for DMA is that the flux of charged
particles from DMA is strongly reduced as compared to the fluxes in isotropic propagation models, which
have been used to claim that the EGRET excess cannot originate from DMA, since this would
overproduce antiprotons \cite{bergstrom_pb}.  However, isotropic propagation models are
ruled out by the observation of convection (ROSAT) \cite{rosat,breitschwerdt_nature} as well as the observations from INTEGRAL
concerning the large bulge/disk ratio for the positron 511 keV annihilation line \cite{prantzos,gebauer}.

The antiproton flux tends to be on the low side in conventional propagation models, both isotropic and anisotropic, as long as  one assumes the local observed shape of the proton spectra is valid for the whole Galaxy, which is a reasonable assumption given the fact that the diffusion of protons is fast compared with the energy losses. In such models DMA simultaneously increases the
gamma rays at energies above 1 GeV and the antiprotons above 0.1 GeV, in excellent agreement
with the data.

The increase in electrons/positrons at high energies by ATIC and positrons by PAMELA may be related, but the fact that ATIC also sees an increase in protons (see \ref{f3}) makes a DMA interpretation unlikely, which agrees with the fact  that positrons are hardly affected by DMA, as shown in Fig. \ref{f4}.
The difference between antiprotons and positrons stems from the fact that positrons have almost no threshold effect, while antiprotons can only be produced by protons above 7 GeV, since
at least 3 additional protons have to appear in the final state to comply with baryon number conservation.
Therefore, the interpretation of the rise in the positron fraction as  signal of  neutralino annihilation is incompatible with the EGRET excess of diffuse gamma rays in an anisotropic propagation model
(see middle panel of Fig. \ref{f4}), while  the flux of antiprotons
  is most easily explained by including a contribution from DMA (in agreement with the EGRET excess).
To fully understand the rise in the positron fraction would require much better data for spectra of electrons, protons and positrons separately instead of only knowing ratios.

\smallskip
 I wish to thank my close collaborators Iris Gebauer, Markus Weber, Dmitri Kazakov and Valery Zhukov for helpful
discussions.
 This work was supported by the BMBF (Bundesministerium f\"ur Bildung und Forschung) via the DLR
(Deutsches Zentrum f\"ur Luft- und Raumfahrt).

\end{document}

%% file: config.tex
\renewcommand\topfraction{.95}
\renewcommand\bottomfraction{.95}
\renewcommand{\floatpagefraction}{0.9}
\renewcommand{\textfraction}{0.05}

\newcommand{\unity}{\mathbf{1}}
\newcommand{\varunity}{\mbox{\rmfamily 1\hspace{-0.25em}l}}

\newcommand{\Dc}{\mathcal{D}}
\newcommand{\Hc}{\mathcal{H}}
\newcommand{\Lc}{\mathcal{L}}
\newcommand{\Oc}{\mathcal{O}}
\newcommand{\Uc}{\mathcal{U}}

\newcommand{\Rep}{\mbox{Re}}
\newcommand{\Imp}{\mbox{Im}}

\newcommand{\Rm}{\mathbb{R}}

\newcommand{\mfbox}[1]{\fbox{$\displaystyle #1$}}


\newlength{\dslashwidth}
\newcommand{\dslash}[1]{\settowidth{\dslashwidth}{$\diagup$}\mbox{%
\hspace{0.5\dslashwidth}\makebox[0pt]{$#1$}\hspace{-0.5\dslashwidth}%
$\diagup$}}

\newcommand{\bsg}{\ensuremath{b\to X_s\gamma}}
\newcommand{\ch}{\ensuremath{\tilde{\chi}^{\pm}}}
\newcommand{\neu}{\ensuremath{\tilde{\chi}^{0}}}
\newcommand{\sinw}{\ensuremath{\sin^2\theta_W}}
\newcommand{\Mgut}{\ensuremath{M_{\mbox{\scriptsize{GUT}}}}}
\newcommand{\agut}{\ensuremath{\alpha_{\mbox{\scriptsize{GUT}}}}}
\newcommand{\tb}{\ensuremath{\tan\beta}}

\def\aii{\alpha_i^{-1}}
\def\rZ{{\rm Z}}
\def\rW{{\rm W}}
\def\rG{{\rm GUT}}
\def\rS{{\rm SUSY}}
\def\rH{{\rm Higgs}}
\def\rF{{\rm Fam}}
\def\MG{M_\rG}
\newcommand{\mc}{Monte Carlo }
\newcommand{\mcs}{Monte Carlos }
\newcommand{\brem}{brems\-strah\-lung }
\newcommand{\bq}{\begin{equation}}
\newcommand{\eq}{\end{equation}}
\newcommand{\ba}{\begin{array}}
\newcommand{\ea}{\end{array}}
\newcommand{\bqa}{\begin{eqnarray}}
\newcommand{\eqa}{\end{eqnarray}}
\newcommand{\nn}{\nonumber \\}
\newcommand{\mpmm}{\mu^{+}\mu^{-}}
\newcommand{\tptm}{\tau^{+}\tau^{-}}
\newcommand{\sq}{^{2}}
\newcommand{\lnf}{{\ifmmode \Lambda^{(N_f)} \else $\Lambda^{(N_f)}$\fi}}
\newcommand{\ms}{{\ifmmode \overline{MS} \else $\overline{MS}$\fi}}
\newcommand{\dr}{{\ifmmode \overline{DR} \else $\overline{DR}$\fi}}
\newcommand{\lms}{{\ifmmode \Lambda^{(5)}_{\overline{MS}} \else $\Lambda^{(5)}_{\overline{MS}}$\fi}}
\newcommand{\lam}{{\ifmmode \Lambda \else $\Lambda$\fi}}
\newcommand{\gev}{{\ifmmode {\rm GeV} \else ${\rm GeV}$\fi}}
\newcommand{\gevc}{{\ifmmode {\rm GeV/c^2} \else ${\rm GeV/c^2}$\fi}}
\newcommand{\tev}{{\ifmmode {\rm TeV} \else ${\rm TeV}$\fi}}
\newcommand{\tevc}{{\ifmmode {\rm TeV/c^2} \else ${\rm TeV/c^2}$\fi}}
\newcommand{\lp}{{\ifmmode L^+  \else $L^+$\fi}}
\newcommand{\lm}{{\ifmmode L^-  \else $L^-$\fi}}
\newcommand{\mlp}{{\ifmmode M(L^-) \else $M(L^-)$\fi}}
\newcommand{\mlz}{{\ifmmode M(L^0) \else $M(L^0)$\fi}}
\newcommand{\lz}{{\ifmmode L^0 \else $L^0$\fi}}
\newcommand{\ev}{{\ifmmode GeV/c^2 \else $GeV/c^2$\fi}}
\newcommand{\tri}{{\ifmmode \triangleup \else $\triangleup$\fi}}
\newcommand{\unl}{{\ifmmode U_{lL^0} \else $U_{lL^0}$\fi}}\newcommand{\gL}{{\ifmmode g_L \else $g_{L}$\fi}}
\newcommand{\gR}{{\ifmmode g_R  \else $g_{R}$\fi}}
\newcommand{\gumu}{{\ifmmode \gamma^{\mu} \else $\gamma^{\mu}$\fi}}
\newcommand{\gunu}{{\ifmmode \gamma^{\nu} \else $\gamma^{\nu}$\fi}}
\newcommand{\gdmu}{{\ifmmode \gamma_{\mu} \else $\gamma_{\mu}$\fi}}
\newcommand{\gdnu}{{\ifmmode \gamma_{\nu} \else $\gamma_{\nu}$\fi}}
\newcommand{\stw}{{\ifmmode\sin^2\theta_W \else $\sin^{2}\theta_{W}$ \fi}}
\newcommand{\sws}{{\ifmmode \;\sin^2\theta_W  \else $\;\sin^{2}\theta_{W}$ \fi}}
\newcommand{\cws}{{\ifmmode \;\cos^2\theta_W  \else $\;\cos^{2}\theta_{W}$ \fi}}
\newcommand{\sw}{{\ifmmode \;\sin\theta_W  \else $\sin\theta_{W}$ \fi}}
\newcommand{\cw}{{\ifmmode \;\cos\theta_W  \else $\;\cos\theta_{W}$ \fi}}
\newcommand{\tw}{{\ifmmode \;\tan\theta_W  \else $\;\tan\theta_{W}$ \fi}}
\newcommand{\qq}{{\ifmmode q\overline{q} \else $q\overline{q}$\fi}}
\newcommand{\lR}{{\ifmmode l_R \else $l_R$\fi}}
\newcommand{\lL}{{\ifmmode l_L \else $l_L$\fi}}
\newcommand{\nt}{{\ifmmode \nu_{\tau} \else $\nu_{\tau}$\fi}}
\newcommand{\nuR}{{\ifmmode \nu_R  \else $\nu_R$\fi}}
\newcommand{\nuL}{{\ifmmode \nu_L  \else $\nu_L$\fi}}
\newcommand{\qR}{{\ifmmode g_R  \else $q_R$\fi}}
\newcommand{\qL}{{\ifmmode q_L  \else $q_L$\fi}}
\newcommand{\qRp}{{\ifmmode q_R'  \else $q_{R}$'\fi}}
\newcommand{\qLp}{{\ifmmode q_L'  \else $q_{L}$'\fi}}
\newcommand{\est}{{\ifmmode e^{\bf \ast} \else $e^{\bf \ast}$\fi}}
\newcommand{\lst}{{\ifmmode l^{\bf \ast} \else $l^{\bf \ast}$\fi}}
\newcommand{\must}{{\ifmmode \mu^{\bf \ast} \else $\mu^{\bf \ast}$\fi}}
\newcommand{\taust}{{\ifmmode \tau^{\bf \ast} \else $\tau^{\bf \ast}$ \fi}}
\newcommand{\pperp}{{\ifmmode p_t  \else $p_t$\fi}}
\newcommand{\et}{{\ifmmode E_t  \else $E_t$\fi}}
\newcommand{\xt}{{\ifmmode x_t  \else $x_t$\fi}}
\newcommand{\smumu}{{\ifmmode \sigma_{\mu\mu}  \else $\sigma_{\mu\mu}$ \fi}}
\newcommand{\eg}{{\ifmmode e\gamma  \else $e\gamma$\fi}}
\newcommand{\epem}{{\ifmmode e^+e^-  \else $e^+e^-$\fi}}
\newcommand{\lplm}{{\ifmmode L^+L^-  \else $L^+L^-$\fi}}
\newcommand{\pp}{{\ifmmode p\overline p  \else $p\overline p$\fi}}
\newcommand{\llz}{{\ifmmode L^0\overline{L}^0 \else $L^0\overline{L}^0$\fi}}
\newcommand{\epemt}{{\ifmmode e^+e^- \to  \else $e^+e^- \to$\fi}}
\newcommand{\eb}{{\ifmmode E_{beam}  \else $E_{beam}$\fi}}
\newcommand{\ip}{{\ifmmode pb^{-1}  \else $pb^{-1}$\fi}}
\newcommand{\upm}{{\ifmmode ^{\pm}  \else $^{\pm}$\fi}}
\newcommand{\de}{{\ifmmode ^{\circ}  \else $^{\circ}$ \fi}}
\newcommand{\appr}{{\ifmmode \sim \else $\sim$ \fi}}
\newcommand{\corresp}{{\ifmmode \stackrel{\wedge}{=} \else $\stackrel{\wedge}{=}$ \fi}}
\newcommand{\sqrts}{{\ifmmode \sqrt{s} \else $\sqrt{s}$\fi}}
\newcommand{\zz}{{\ifmmode Z^0  \else $Z^0$\fi}}
\newcommand{\mz}{{\ifmmode M_{Z}  \else $M_{Z}$\fi}}
\newcommand{\mzs}{{\ifmmode M_{Z}^2  \else $M_{Z}^2$\fi}}
\newcommand{\mw}{{\ifmmode M_{W}  \else $M_{W}$\fi}}
\newcommand{\mws}{{\ifmmode M_{W}^2  \else $M_{W}^2$\fi}}
\newcommand{\mh}{{\ifmmode M_{Higgs}  \else $M_{Higgs}$\fi}}
\newcommand{\gt}{{\ifmmode \Gamma_{tot} \else $\Gamma_{tot}$\fi}}
\newcommand{\msusy}{{\ifmmode M_{SUSY}  \else $M_{SUSY}$\fi}}
\newcommand{\msusys}{{\ifmmode M_{SUSY}^2  \else $M_{SUSY}^2$\fi}}
\newcommand{\su}{{\ifmmode SU(3)_C\otimes\- SU(2)_L\otimes\- U(1)_Y \else $SU(3)_C\otimes SU(2)_L\otimes U(1)_Y$\fi}}
\newcommand{\suthree}{{\ifmmode SU(3)_C  \else $SU(3)_C$\fi}}
\newcommand{\sutwo}{{\ifmmode  SU(2)_L\otimes U(1)_Y \else $SU(2)_L\otimes U(1)_Y$\fi}}
\newcommand{\taup} {{\ifmmode \tau_{proton} \else $\tau_{proton}$\fi}}
\newcommand{\as}{{\ifmmode \alpha_{s}  \else $\alpha_{s}$\fi}}
\newcommand{\mgut}{{\ifmmode M_{GUT}  \else $M_{GUT}$\fi}}
\newcommand{\mguts}{{\ifmmode M_{GUT}^2  \else $M_{GUT}^2$\fi}}
\newcommand{\mze} {{\ifmmode m_0        \else $m_0$\fi}}
\newcommand{\mha}{{\ifmmode m_{1/2}    \else $m_{1/2}$\fi}}
\newcommand{\mb} {{\ifmmode m_{b}    \else $m_{b}$\fi}}
\newcommand{\mt} {{\ifmmode m_{t}    \else $m_{t}$\fi}}
\newcommand{\mts} {{\ifmmode m_{t}^2    \else $m_{t}^2$\fi}}
\newcommand {\rb}[1]{\raisebox{1.5ex}[-1.5ex]{#1}}
\newcommand{\mtau}{{\ifmmode m_{\tau}  \else $m_{\tau}$\fi}}
\newcommand{\dpp}{{\ifmmode \delta_{pert} \else $\delta_{pert}$\fi}}
\newcommand{\dnp}{{\ifmmode\delta_{non-pert}\else$\delta_{non-pert}$\fi}}
\newcommand{\dew}{{\ifmmode \delta_{\rm EW}\else $\delta_{\rm EW}$\fi}}
\newcommand{\rt}{{\ifmmode R_{\tau}  \else $R_{\tau} $\fi}}
\newcommand{\rz}{{\ifmmode R_{Z}  \else $R_{Z} $\fi}}
\newcommand{\into}{\rightarrow}
\newcommand{\SM}{Standard Model}
\newcommand{\swb}{{\ifmmode \sin^2\theta_{\overline{MS}} \else $\sin^2\theta_{\overline{MS}}$\fi}}
\newcommand{\cwb}{{\ifmmode \cos^2\theta_{\overline{MS}} \else $\cos^2\theta_{\overline{MS}}$\fi}}
\newcommand{\ttbs}{\char'134}
\newcommand{\besg}{$b  \to  X_s \gamma~ $}
\newcommand{\mzero}{\rm m_0}
\newcommand{\mhalf}{\rm m_{1/2}}